\documentclass[12pts,amsfonts]{article}
\usepackage{amsfonts}
\renewcommand{\Im}{\mbox{Im }}

\newcommand{\ra}{\rightarrow}
\newcommand{\bra}{\langle} \newcommand{\ket}{\rangle}
\newcommand{\be}{\begin{equation}}
\newcommand{\ee}{\end{equation}}
\newcommand{\bea}{\begin{eqnarray}}
\newcommand{\eea}{\end{eqnarray}}
\newcommand{\eps}{\epsilon}

\newcommand{\ffi}{\varphi}

\newcommand{\ep}{\qquad {\vrule height 10pt width 8pt depth 0pt}}

\newcommand{\grintl}{[\kern-.18em [}
\newcommand{\grintr}{]\kern-.18em ]}

\newtheorem{prop}{Proposition}[section]
\newtheorem{thm}{Theorem}[section]
\newtheorem{cor}{Corollary}[section]

\def\smallR{\hbox{\scriptsize I\kern-.23em{R}}}
\def\R{\mathbb R}
\def\C{\mathbb C}
\def\un{\mathbb I}
\def\0{\hbox{$\mit I$\kern-.70em$\mit O$}}
\def\r{I\kern-.277em R}

\def\N{\mbox{\bf N}}

\begin{document}

\title{Abstract adiabatic charge pumping}
\author{A. Joye,\footnote{Partially supported by the Agence Nationale de la Recherche, grant ANR-09-BLAN-0098-01}\\
Institut Fourier, Universit\'e de Grenoble 1\\BP 74, 38402 Saint-Martin-d'H\`eres Cedex, France\\
\\ V. Brosco, \\
Dipartimento di Fisica, Universit\'a di Roma ``La Sapienza''\\
P.le A. Moro, 2 00185 Roma, Italy\\
\& ISC-CNR, via dei Taurini, 19 00185 Roma, Italy \\
\\ F. Hekking, \\
LPMMC, CNRS \& Universit\'e  de Grenoble 1\\ BP 166,
        38042 Grenoble CEDEX 9, France }
\maketitle
\abstract{
This paper is devoted to the analysis of an abstract formula
describing quantum adiabatic charge pumping in a general
context. We consider closed systems characterized by a slowly
varying time-dependent Hamiltonian depending on an external parameter
$\alpha$.  The current operator, defined as the derivative of the
Hamiltonian with respect to  $\alpha$, once integrated over some time interval, gives rise to a
charge pumped through the system over that time span.  We determine the first two leading
terms in the adiabatic parameter of this pumped charge under the usual gap
hypothesis. In particular, in  case the  Hamiltonian is time periodic and has discrete non-degenerate spectrum, the charge pumped over a period is given to leading order by the derivative
with respect to $\alpha$ of the corresponding dynamical and geometric phases.
}

\section{Introduction}

\subsection{Motivation}

Many physical systems of interest can be described by means of a  time
dependent Hamiltonian reflecting the action of external agents on the
system or taking into account the variations of its environment, in
an effective manner. In such generality, little can  be said about the
evolution of the system. However, when the Hamiltonian is a slowly
varying function of time with respect to some fixed relevant time
scale of the system, the adiabatic theorem of Quantum Mechanics
provides a very useful tool to describe the evolution in an
approximate way under certain hypotheses, see e.g \cite{bf, mes}. The mathematical circumstances under which an adiabatic theorem of Quantum Mechanics can be proven are diverse, starting with the well known gap hypothesis in the spectrum of the Hamiltonian, see e.g. \cite{k1, n1, asy}, which will be enough for our  purposes. Note, however, that higher order approximations \cite{n2, n3, jp2}, generalizations to situations where the gap assumption fails
\cite{bornemann, ae} or where self-adjointness does not hold  \cite{lidar, joye} or to a space-time setting \cite{t} have been carried out over the years; see also the review \cite{hjbham}  and references therein.

In several physical systems, the Hamiltonian also depends on some external parameter, $\alpha\in \mathbb R$, associated with an instantaneous charge {\it current}. For instance, this is the case for models used in the study of the Quantum Hall effect where the electric current is monitored by a magnetic flux, $\alpha$, through the sample which appears as a parameter in the Hamiltonian $H_\alpha(s)$. The instantaneous current operator is given by $\partial_\alpha H_\alpha(s)$ see e.g. \cite{nt, asy}. A similar phenomenon holds in models of mesoscopic physics where the current across the device they describe is driven by a phase difference. Examples are superconducting systems used as Cooper pair pumps~\cite{pekola,aunola03,fhp,governale,mottonen06,mottonen08,bfhj}.

Accordingly, for an initial state $\psi$, the charge pumped through the system in that state between time $0$ and $\tau$  is given  by
$$\bra Q_\alpha(\tau)\ket_\psi:=\int_0^\tau \bra U_\alpha(\tau ')\psi | ( \partial_\alpha H_\alpha(\tau')) U_\alpha(\tau')\psi\ket d\tau',$$ where $U_\alpha(\tau)$ is the evolution operator. It is often true in applications that the time variation of the hamiltonian is periodic, in which case one considers the charge transported over a time period. For a slowly varying Hamiltonian $H_\alpha$ with an isolated part of spectrum and an initial states $\psi$ in the corresponding spectral subspace, the pumped charged $\bra Q_\alpha(\tau)\ket_\psi$ can be computed by making use of an adiabatic approximation. This yields the starting point of the analysis of the geometrical and topological properties of this quantity. In particular, when the dependence of the hamiltonian is periodic in both the time and parameter $\alpha$, it is well known that  the pumped charge over a cycle suitably  {\em averaged}  over $\alpha$ is quantized and related to topological indices associated with the spectral projector, see e.g. \cite{nt, as, asy}. However, in certain mesoscopic devices, the observed charge transfers do correspond to $\bra Q_\alpha(\tau)\ket_\psi$ for fixed values of $\alpha$ \cite{bfhj}, which is also known do differ significantly from the average value over $\alpha$, \cite{asy}.

Note that in contrast to other mathematical studies, see e.g. \cite{aegss}, the time-dependent adiabatic pumps we consider are not open systems in the following sense: their dynamics for frozen times is not characterized by an explicit scattering matrix between infinite reservoirs. Therefore, no formula is available to determine the charge transport in terms of matrix elements of the instantaneous scattering matrix \cite{brow, btp} .
\medskip

The purpose of this paper is to provide a mathematical derivation of the adiabatic pumped charge through the system in an abstract setting that covers the physical situations described above.

We focus on the derivation of a controlled approximate expression for $\bra Q_\alpha(\tau)\ket_\psi$, the charge pumped through the system over a time span $[0,\tau]$  for fixed values of $\alpha$ up to vanishing corrections in the adiabatic regime. We work in a quite general framework, with arbitrary dependence of the Hamiltonian on time and $\alpha$, assuming only the existence of an isolated part of spectrum in $H_\alpha(\tau)$. This allows to recover as particular cases the time-periodic situations and, for an isolated eigenvalue in the spectrum, expressions for the charge pumped over a period used in several contexts, involving the geometric phase or its non-abelian  version \cite{bfhj}.

Moreover, we hope the present analysis can provide an introduction to the mathematical methods used in the analysis of adiabatic phenomena beyond the matrix case.

\medskip
\subsection{Setup and Results}

Let us describe more precisely the mathematical setup and results we prove in the next Section.  The adiabatic regime is characterized by a dimensionless time scale $1/\eps$, which is assumed to be long, i.e.  $\eps \ll 1$. The Hamiltonian is assumed to depend on $\eps$ as follows: $H_\alpha=H_\alpha(\eps \tau)$. This makes  the Hamiltonian slowly varying since it takes a time of order $1/\eps$ for it to change significantly. Introducing the rescaled time variable $t=\eps \tau$, of order one, the corresponding evolution operator $U_\alpha$ also depends on the small parameter $\eps$ and satisfies
$$i\eps \partial_t U_\alpha(t)=H_\alpha(t)U_\alpha(t), \ \ \ U_\alpha(0)={\mathbb I}.$$
Accordingly, for any initial state $\psi$, the charge pumped over a time interval $[0,\tau=t/\eps]$ now reads
$$
\bra Q_\alpha(t) \ket_\psi=\frac{1}{\eps}\int_0^t\bra
U_\alpha(s)\psi |
(\partial_\alpha H_\alpha(s))U_\alpha(s)\psi \ket ds,
$$
which is of order $1/\eps$. From here on, s is consistently the rescaled dummy integration time variable. In order to capture the leading term, up to a vanishing error in the adiabatic limit $\eps\ra 0$, it is necessary to compute the next to leading order approximation of the evolution operator in $\eps$. We do this in Theorem \ref{sec}, under the sole gap hypothesis, after having recalled the usual leading order adiabatic approximation in Theorem \ref{adiab}.  Then we focus on initial conditions that belong to the spectral projector corresponding to the isolated part of spectrum by the gap. The corresponding general expression for the charge operator is provided in Proposition \ref{prop}. A far more tractable expression is obtained for an isolated eigenvalue, of arbitrary degeneracy, as  Corollary \ref{isolev}. In case the Hamiltonian is time periodic of $t$ period one, we recover in Corollary \ref{evper} the expressions used in \cite{bfhj}. The simplest instance being for a simple isolated eigenvalue $E_\alpha(t)$, such that $H_\alpha(t)\psi(t)=E_\alpha(t)\psi(t)$. The charge pumped over a cycle for the initial condition $\psi=\psi(0)$ then  reads
$$
\bra Q_\alpha(1) \ket_\psi=\frac{1}{\eps}\int_0^1
\partial_\alpha E_\alpha(s)ds + \partial_\alpha \beta_\alpha+O(\eps).
$$
Here the first term is the usual dynamic contribution, whereas the second one is of geometric nature~\cite{simon,wilczek}, $\beta_\alpha$ being the familiar geometric phase \cite{be,bohm}. In case the eigenvalue is degenerate, the matrix valued non-abelian generalization of this quantity, $B(1)$, to be used is of course also described in Corollary \ref{isolev}.

The last Section of the paper is devoted to the study of a family of examples used in the physical application described in \cite{bfhj}. The Hamiltonians considered display a permanently degenerate isolated eigenvalue for which explicit computations of $B(1)$ can be performed. In case the dimension of the corresponding degenerate eigenspace is two, which corresponds to the applications considered in \cite{bfhj}, we also provide a geometric interpretation of $B(1)$.

\section{Analysis}

In this section, we provide the abstract rigorous mathematical analysis
behind the formulas used in study of the physical phenomenon of adiabatic
charge pumping. We feel such a rigorous analysis is useful because of the presence
of a variety of formulations of geometric adiabatic charge pumping in the
litterature which have similar features. This allows to make clear under which
hypotheses we work. Also, we believe the analysis is interesting in
itself because it applies under very general conditions and it might be of use in
different frameworks.\\

Let ${\cal H}$ be a separable Hilbert space and $H_\alpha(t)$ be a
bounded self-adjoint operator on ${\cal H}$. In order not to obscure the analysis
by side issues regarding the technical difficulties related to the use of
unbounded operators, we stick to the bounded case. For the same reason, we assume the parameter $\alpha$ is real valued. We work under the \\

{\bf Regularity assumption:}
The map $(t,\alpha)\mapsto H_\alpha(t)$ is $C^3$ in the norm sense,
as a bounded operator
valued function, with $(t,\alpha)\in [0,1]\times [0,1]$. \\

Let
$U_\alpha(t)$ be the solution to
\be\label{se}
i\eps \dot U_\alpha(t)=H_\alpha(t)U_\alpha(t), \ \ U_\alpha(0)=\mathbb I,
\ee
where we rescaled time for convenience.
Since ${\cal L}({\cal H})$ is a Banach space, it follows from the  general theory
of differential equations, see e.g. \cite{lang} Ch. VI, that
the solution to (\ref{se}) is as regular in $(t,\alpha)$ as the Hamiltonian is, i.e.
\be
(t,\alpha) \mapsto U_\alpha(t) \in C^3([0,1]\times [0,1], {\cal L}({\cal H})).
\ee

The variable $\alpha$ is a parameter whose variations monitor fluxes
or currents in the time dependent physical device described by $H_\alpha(t)$.
The {\em current} operator being {\em defined} as $\partial_\alpha H_\alpha (t)$, if
$\psi\in  {\cal H}$ is a normalized initial state the average charge
pumped by the system between the physical times $0$ and $t/\eps$ is equal to
\be
\bra Q_\alpha(t) \ket_\psi=\frac{1}{\eps}\int_0^t\bra
U_\alpha(s)\psi |
(\partial_\alpha H_\alpha(s))U_\alpha(s)\psi \ket ds.
\ee

This leads us to study the charge operator $Q_\alpha(t)$, whose matrix elements carry the
physical interpretation,  defined by
\be\label{defq}
Q_\alpha(t)=\frac{1}{\eps}\int_0^t
U_\alpha^{-1}(s)
(\partial_\alpha H_\alpha(s))U_\alpha(s) ds.
\ee
The geometrical properties of another {\em average } charge operator in the adiabatic  limit
have been investigated  in \cite{as} under the assumption that
$(t,\alpha)\mapsto H_\alpha(t)$ is periodic in both variables. The average there is
taken both over the period of the time dependent Hamiltonian and over the flux
variable $\alpha$. By contrast, we  analyze the adiabatic behaviour of
the charge operator for fixed values of $\alpha$ and $t$. In case of  a time
periodic Hamiltonian, the geometric content of the charge operator over a period
is elucidated.

\vspace{.3cm}

As already observed in \cite{asy} for example, an alternative exact expression for $Q_\alpha(t)$
reads as follows:
\be\label{alternate}
Q_\alpha(t)=iU_\alpha^{-1}(t)\partial_\alpha U_\alpha(t).
\ee
It is a consequence of the regularity of $U_\alpha(t)$ and the computation
\bea
i\eps \partial_t \left(U_\alpha^{-1}(t)\partial_\alpha U_\alpha(t)\right)
&=&-U_\alpha^{-1}(t)H_\alpha(t)\partial_\alpha U_\alpha(t)\nonumber\\
&+&U_\alpha^{-1}(t)\partial_\alpha\left(H_\alpha(t) U_\alpha(t)\right)
\nonumber\\
&=&U_\alpha^{-1}(t)\left( \partial_\alpha H_\alpha(t)\right) U_\alpha(t),
\eea
with $\partial_\alpha U_\alpha(0)=0$.
Note that we also deduce from (\ref{alternate}) the general estimate
\be
\partial_\alpha U_\alpha(t)=O(t/\eps),
\ee
which is sharp in the scalar case.

\vspace{.3cm}

We proceed by considering the adiabatic limit $\eps\ra 0$ under
the familar \vspace{.3cm}

{\bf Gap hypothesis}:
Assume the spectrum $\sigma(H_\alpha(t))$ consists in two
disjoint  parts $\sigma(H_\alpha(t))=\sigma_\alpha(t) \cup \overline{\sigma_\alpha}(t)$
such that
\be\label{gap}
\inf_{(t,\alpha)\in [0,1]^2}
\mbox{dist} (\sigma_\alpha(t),\overline{\sigma_\alpha}(t))=g>0.
\ee

Let $P_\alpha(t)$ be the spectral projector of $H_\alpha(t)$
associated with $\sigma_\alpha(t)$ by means of the Riesz formula
\be\label{riesz}
P_\alpha(t)=-\frac{1}{2\pi i}\int_\gamma (H_\alpha(t)-z)^{-1}\, dz \, ,
\ee
where $\gamma $ is a loop in the complex plane encircling $\sigma_\alpha(t)$
only, which is locally independent of $(t, \alpha )$
and let
$\overline{P_\alpha}(t)=\mathbb I -P_\alpha(t)$ be its complement. These projectors
are as regular as the Halmiltonian is and moreover satisfy for any $(t,\alpha)$
\be\label{ppp}
P_\alpha(t) \dot P_\alpha(t) P_\alpha(t)=P_\alpha(t) (\partial_\alpha P_\alpha(t))
P_\alpha(t)\equiv 0,
\ee
as easily  seen by differentiating the identity $P_\alpha(t)^2=P_\alpha(t)$.

We introduce two unitary operators whose product will approach the true
evolution  in the adiabatic limit.
Let $W_\alpha(t)$ and $\Phi_\alpha(t)$ be defined by
\bea\label{defwdefp}
i\dot W_\alpha(t)&=&K_\alpha(t)W_\alpha(t),  \ \ W_\alpha(0)=\mathbb I,\\
i\eps \dot\Phi_\alpha(t)&=&W_\alpha^{-1}(t)H_\alpha(t)W_\alpha(t) \Phi_\alpha(t),
\ \ \Phi_\alpha(0)=\mathbb I,
\eea
where
\be
K_\alpha(t)=i[\dot P_\alpha(t),P_\alpha(t)].
\ee
It is a classical fact that the following relations hold,
\bea\label{clas}
 W_\alpha(t)P_\alpha(0)&=&P_\alpha(t)W_\alpha(t)\\ \label{clas2}
 \Phi_\alpha(t)P_\alpha(0)&=&P_\alpha(0)\Phi_\alpha(t).
\eea
The first identity is proven by showing that both operators satisfy the same
differential equation with same initial condition, exploiting the relation
\be\label{dk}
P_\alpha(t)K_\alpha(t)P_\alpha(t)=0,
\ee
which is a consequence of (\ref{ppp}), see e.g. \cite{k1}. The second identity follows from the fact that,
by construction, the generator of $\Phi_\alpha(t)$ commutes with $P_\alpha(0)$.\\

We have the
\begin{thm}[Adiabatic Theorem] \label{adiab} Assuming the Regularity and
Gap hypotheses above, we have
for any $(t,\alpha)\in [0,1]\times [0,1]$,
\be
U_\alpha(t)= W_\alpha(t)\Phi_\alpha(t)+O(\eps),
\ee
where the error term is uniform in $\alpha \in [0,1]$.
\end{thm}
{\bf Remarks:}\\
i) As stated, the Theorem  dates back  to \cite{n1}, \cite{n2} and was generalized
by \cite{asy} to the unbounded case. The fact that the error term is uniform
in $\alpha\in[0,1]$ is a straighforward consequence of our Regularity Assumtions and of the Gap hypothesis which
is uniform in $\alpha\in[0,1]$.\\
ii) In case $\sigma_\alpha(t)$ consists  of a single eigenvalue $E_\alpha(t)$,
the Theorem says
\be
U_\alpha(t)P_\alpha(0)=e^{-\frac{i}{\eps}\int_0^tE_\alpha(s)ds}P_\alpha(t)W_\alpha(t)
+O(\eps).
\ee
This is the statement proven by Kato in \cite{k1}.\\
ii) Further assuming $E_\alpha(t)$ is nondegenerate and
the time-dependent Hamiltonian is periodic in time, of period $1$,
we get the {\em geometric} or Berry phase out of this formula as follows:

Let $\ffi_\alpha(0)=P_\alpha(0)\ffi_\alpha(0)$, be a normalized eigenvector
associated with $E_\alpha(0)$. Then, property (\ref{clas}) implies
that $\forall (\alpha,t)\in [0,1]^2$
\be\label{evt}
\ffi_\alpha(t)=W_\alpha(t)\ffi_\alpha(0) \ \ \mbox{satisfies} \ \
H_\alpha(t)\ffi_\alpha(t)=E_\alpha(t)\ffi_\alpha(t).
\ee
and (\ref{defwdefp}) together with (\ref{dk}) yield
\be\label{dfi}
\bra \ffi_\alpha(t) |\dot{\ffi}_\alpha(t)\ket =0.
\ee
By periodicity, $\ffi_\alpha(0)$ and $\ffi_\alpha(1)$ differ by a phase only
\be\label{berry}
\ffi_\alpha(1)=e^{-i\beta_\alpha}\ffi_\alpha(0),
\ee
where $\beta_\alpha$ is the geometric or Berry phase.
Therefore we finally get
\be
U_\alpha(1)\ffi_\alpha(0)=e^{-\frac{i}{\eps}\int_0^1E_\alpha(s)ds}e^{-i\beta_\alpha}
\ffi_\alpha(0)+O(\eps).
\ee
This shows concretely that
the operator $W_\alpha(t)$ carries the geometrical content of the adiabatic Theorem.
Note finally that if
$t\mapsto \psi_\alpha(t)$ is another choice of instantaneous normalized
eigenvector associated with $E_\alpha(t)$ which is periodic in time,
these vectors differ from $\ffi_\alpha(t)$ by a phase for any time
\be
\ffi_\alpha(t)=e^{-ib_\alpha(t)}\psi_\alpha(t).
\ee
Using (\ref{dfi}), we get an explicit expression for $b$
\be
b_\alpha(t)=b_\alpha(0)-i\int_0^t \bra \psi_\alpha(s) | \dot{\psi}_\alpha(s)\ket \, ds.
\ee
Therefore
$
\ffi_\alpha(1)=
e^{-\int_0^1\,\bra \psi_\alpha(s) | \dot{\psi}_\alpha(s)\ket ds}\ffi_\alpha(0)
$
and
\be\label{exbe}
\beta_\alpha=-i\int_0^1\,\bra \psi_\alpha(s) | \dot{\psi}_\alpha(s)\ket ds.
\ee
\vspace{.3cm}

Coming back to the charge operator, we see that in order to estimate
$Q_\alpha(t)$ up to errors of order $O(\eps)$ vanishing in the adiabatic limit,
we need to control the evolution to order $O(\eps^2)$, see (\ref{defq}). This can be
achieved as follows, see e.g. \cite{n2},\cite{asy}, \cite{jp2},  \cite{n3},  \dots.

Let
\be
H^{(1)}_\alpha(t)=H_\alpha(t)-\eps K_\alpha(t),
\ee
which satisfies the gap assumption (\ref{gap}) for $\eps $ small enough since $K_\alpha(t)$
is bounded i.e.
\be
\sigma(H^{(1)}_\alpha(t))=\sigma^{(1)}_\alpha(t)\cup \overline{\sigma^{(1)}_\alpha}(t).
\ee
Hence we can define the
corresponding spectral projectors $P^{(1)}_\alpha(t)$ by (\ref{riesz}) and
$\overline{P^{(1)}_\alpha}(t)=\mathbb I- P^{(1)}_\alpha(t)$. By perturbation
theory we have for $\eps$ small enough
\be\label{perres}
(H^{(1)}_\alpha(t)-z)^{-1}=(H_\alpha(t)-z)^{-1}+
(H_\alpha(t)-z)^{-1}\eps K_\alpha(t)(H_\alpha(t)-z)^{-1}+O(\eps^2)
\ee
where the remainder term is $C^2$ in $(t,\alpha)$ and of order $\eps^2$, as
a uniformly convergent Neuman series.  The same is true for the
perturbed projector expressed using (\ref{riesz}) with the same path
$\gamma$ for $\eps$ small enough,
\be\label{perpro}
P^{(1)}_\alpha(t)=P_\alpha(t) -\eps {\cal R}_\alpha(K_\alpha)(t)+O(\eps^2),
\ee
where, for any bounded operator $B$,
\be
{\cal R}_\alpha(B)(t)=\frac{1}{2 i \pi }\oint_{\gamma}
(H_\alpha(t)-z)^{-1}B(H_\alpha(t)-z)^{-1}dz
\ee
with $\gamma$ is a loop encircling $\sigma^{(1)}_\alpha(t)$, which can  be chosen
locally independently of $(t,\alpha)$. Let us also note here for future reference that
\be\label{fut}
P_\alpha(t){\cal R}_\alpha(K_\alpha)(t)P_\alpha(t)=0,
\ee
as a consequence of the fact that the resolvent and the spectral projectors of
$H_\alpha(t)$ commute and of (\ref{dk}).
We set
\be
K^{(1)}_\alpha(t)=i[\dot P^{(1)}_\alpha(t),P^{(1)}_\alpha(t)]=K_\alpha(t)+O(\eps),
\ee
where,  again, the error term can be differentiated without harm.
We define $W^{(1)}_\alpha(t)$ and $\Phi^{(1)}_\alpha(t)$
by
\bea
i\dot W^{(1)}_\alpha(t)&=&K^{(1)}_\alpha(t)W^{(1)}_\alpha(t),
\ \ W^{(1)}_\alpha(0)=\mathbb I,\\
i\eps \dot\Phi^{(1)}_\alpha(t)&=&{W^{(1)}_\alpha}^{-1}(t)\left(H^{(1)}_\alpha(t)+\eps
{\cal D}^1_\alpha(K_\alpha(t)\right)W^{(1)}_\alpha(t)
\Phi^{(1)}_\alpha(t),  \\ \nonumber
&& \hspace{6.2cm} \Phi^{(1)}_\alpha(0)=\mathbb I,
\eea
where, for any bounded operator $B$,
${\cal D}^1_\alpha(B)=P^{(1)}_\alpha B P^{(1)}_\alpha +
\overline{P^{(1)}_\alpha} B \overline{P^{(1)}_\alpha}.$
One gets that the relations  equivalent to  (\ref{clas}), (\ref{clas2})
hold with indices
$\phantom{.}^{(1)}$ at the
relevant operators.

The point of this construction is that it gives a
\begin{thm}\label{sec}[Second Order Adiabatic Theorem]
Under the hypotheses of Theorem \ref{adiab}, we have
for any $(t,\alpha)\in [0,1]\times [0,1]$,
\be
U_\alpha(t)= W^{(1)}_\alpha(t)\Phi^{(1)}_\alpha(t)+O(\eps^2),
\ee
where the error term is uniform in $\alpha\in [0,1]$.
\end{thm}
{\bf Remarks:}\\
i)  If the Hamiltonian is regular enough in the $t$ variable,
it is possible to get arbitrary order adiabatic theorems.\\
ii) These adiabatic theorems further yield the perturbative estimate
\be\label{perwp}
W^{(1)}_\alpha(t)\Phi^{(1)}_\alpha(t)=W_\alpha(t)\Phi_\alpha(t)+ O(\eps).
\ee
iii) This result can be found under various guises in \cite{n2},\cite{asy}, \cite{jp2},  \cite{n3},  \dots.  As such, it is stated and proven in \cite{thesis},
Theorem 3.3.1, p. 38 (for $\alpha$ fixed). Again, the uniformity in $\alpha$ of the error term is easily checked.\\

As a consequence of this second Theorem, we have the approximation
\be\label{inter}
Q_\alpha(t)=\frac{1}{\eps}\int_0^t
{\Phi^{(1)}_\alpha}^{-1}(s){W^{(1)}_\alpha}^{-1}(s)
\partial_\alpha H_\alpha(s)W^{(1)}_\alpha(s)\Phi^{(1)}_\alpha(s) ds+O(\eps).
\ee
We point out again that the analog of the formula above with operators
$W_\alpha$ and $\Phi_\alpha$ in place of $W_\alpha^{(1)}$ and
$\Phi_\alpha^{(1)}$ yields an error term of order one, instead of
$O(\eps)$.\\

We are interested in the matrix elements of $Q_\alpha(t)$ with vectors
belonging to the spectral subspace $P_\alpha(0){\cal H}$, so that from now on,
we focus on the operator $P_\alpha(0)Q_\alpha(t)P_\alpha(0)$. The goal is
to express the information in (\ref{inter}) in terms of more familiar quantities,
like dynamical phases and geometric phases, in certain cases.

The next technical result says that we can express $P_\alpha(0)Q_\alpha(t)P_\alpha(0)$
as a leading term (of order $1/\eps$) that corresponds to the replacement of
$U_\alpha(t)$ by its second order adiabatic approximation with
$P^{(1)}_\alpha(0)$ in place of $P_\alpha(0)$ and
$\partial_\alpha H^{(1)}_\alpha(t)$ in place of $\partial_\alpha H_\alpha(t)$
plus a term of order $\eps^0$ which will give rise to the geometric contribution:

\begin{prop}\label{prop}
Under the hypotheses of Theorem \ref{adiab}, we have
for any $(t,\alpha)\in [0,1]\times [0,1]$,
\bea\label{first}
&&P_\alpha(0)Q_\alpha(t)P_\alpha(0)\\
&&=P_\alpha(0)
\left(\frac{1}{\eps}\int_0^t
{\Phi^{(1)}_\alpha}^{-1}(s){W^{(1)}_\alpha}^{-1}(s)
P^{(1)}_\alpha(s) \partial_\alpha H^{(1)}_\alpha(s)
P^{(1)}_\alpha(s) W^{(1)}_\alpha(s) \Phi^{(1)}_\alpha(s)ds\right) P_\alpha(0)\nonumber\\
\nonumber
&&+P_\alpha(0)\left( \int_0^t \Phi_\alpha^{-1}(s)W_\alpha^{-1}(s)
\partial_\alpha K_\alpha(s)
W_\alpha(s)\Phi_\alpha(s) ds\right) P_\alpha(0)+O(\eps).
\eea
\end{prop}
{\bf Proof:} Plugging the relations
\be
P_\alpha(t)=P_\alpha(t)^2=
P_\alpha(t)(P^{(1)}_\alpha(t)-\eps {\cal R}_\alpha(K_\alpha)(t))+O(\eps^2)
\ee
and
\be
\partial_\alpha H_\alpha(t)=\partial_\alpha H^{(1)}_\alpha(t)+
\eps\partial_\alpha K_\alpha(t)
\ee
into the expression (\ref{inter}), and making use of the
properties of $W^{(1)}_\alpha $ and $\Phi^{(1)}_\alpha $ together with (\ref{perwp}),
we get
\bea
&&P_\alpha(0)Q_\alpha(t)P_\alpha(0)\nonumber\\
&&=P_\alpha(0)
\left(\frac{1}{\eps}\int_0^t
{\Phi^{(1)}_\alpha}^{-1}(s){W^{(1)}_\alpha}^{-1}(s)
P^{(1)}_\alpha(s) \partial_\alpha H^{(1)}_\alpha(s)
P^{(1)}_\alpha(s) W^{(1)}_\alpha(s)\Phi^{(1)}_\alpha(s)ds \right) P_\alpha(0)\nonumber\\
&&+P_\alpha(0)\left( \int_0^t \Phi_\alpha^{-1}(s)W_\alpha^{-1}(s)\partial_\alpha K_\alpha(s)
W_\alpha(s)\Phi_\alpha(s) ds\right) P_\alpha(0)\nonumber\\
&&-P_\alpha(0)\left({\cal R}_\alpha(K_\alpha)(0)\, \int_0^t
\Phi_\alpha^{-1}(s)W_\alpha^{-1}(s)
\partial_\alpha H_\alpha(s)
W_\alpha(s)\Phi_\alpha(s) ds\right.\nonumber\\&&\left.+
\int_0^t\Phi_\alpha^{-1}(s)W_\alpha^{-1}(s)
\partial_\alpha H_\alpha(s)
W_\alpha(s)\Phi_\alpha(s) ds\,{\cal R}_\alpha(K_\alpha)(0)\right)P_\alpha(0)
+O(\eps).
\eea
We want to show that the last two terms are actually of order $\eps$, by
integration by parts. We consider the last term only, since the previous
one can be dealt with in a similar fashion. Using property (\ref{fut}),
the integrand of this term is
\be\label{1}
P_\alpha(0)\Phi_\alpha^{-1}(s)W_\alpha^{-1}(s)P_\alpha(s)
\partial_\alpha H_\alpha(s)\overline{P_\alpha}(s)
W_\alpha(s)\Phi_\alpha(s)\overline{P_\alpha}(0).
\ee
Differentiating the following identity with respect to $\alpha$
\be
H_\alpha(s)=P_\alpha(s)H_\alpha(s)P_\alpha(s)
+\overline{P_\alpha}(s)H_\alpha(s)\overline{P_\alpha}(s),
\ee
we get
\bea
P_\alpha(s)
\partial_\alpha H_\alpha(s)\overline{P_\alpha}(s)&=&P_\alpha(s)H_\alpha(s)
\partial_\alpha P_\alpha(s) \overline{P_\alpha}(s)+P_\alpha(s)\partial_\alpha\overline{P_\alpha}(s)H_\alpha(s)\overline{P_\alpha}(s)\nonumber\\
&=&P_\alpha(s)[H_\alpha(s), \partial_\alpha P_\alpha(s)]\overline{P_\alpha}(s).
\eea
Hence, together with (\ref{defwdefp}), we can write
\bea\nonumber
\mbox{(\ref{1})}&=&P_\alpha(0)\Phi_\alpha^{-1}(s)W_\alpha^{-1}(s)[H_\alpha(s), \partial_\alpha P_\alpha(s)]W_\alpha(s)\Phi_\alpha(s)\overline{P_\alpha}(0)\\ \nonumber
&=&P_\alpha(0)\Phi_\alpha^{-1}(s)[W_\alpha^{-1}(s)H_\alpha(s)W_\alpha(s), W_\alpha^{-1}(s)\partial_\alpha P_\alpha(s)W_\alpha(s)] \Phi_\alpha(s)\overline{P_\alpha}(0)\\
\nonumber
&=&-i\eps \partial_s \left(P_\alpha(0)\Phi_\alpha^{-1}(s)W_\alpha^{-1}(s)\partial_\alpha P_\alpha(s)W_\alpha(s)\Phi_\alpha(s)\overline{P_\alpha}(0) \right)\\
& &+ i\eps P_\alpha(0)\Phi_\alpha^{-1}(s)\partial_s\left(  W_\alpha^{-1}(s)\partial_\alpha P_\alpha(s)W_\alpha(s) \right)\Phi_\alpha(s)\overline{P_\alpha}(0).
\eea
Thus, since $W_\alpha(s)$ and $P_\alpha(s)$ are independent of $\eps$, and $\Phi_\alpha(s)$
is unitary, when integrated between $0$ and $t\in [0,1]$, this yield a contribution
of order $\eps$. \ep

\noindent{\bf Remark:}
We can get an alternative expression for the  zero'th order term
by making use of the identity
\bea
 P_\alpha(s)\partial_\alpha K_\alpha(s) P_\alpha(s)&=&i P_\alpha(s)\partial_\alpha [\dot P_\alpha(s), P_\alpha(s)] P_\alpha(s)\\  \nonumber
&=&i P_\alpha(s)[\dot P_\alpha(s), \partial_\alpha P_\alpha(s)] P_\alpha(s).
\eea

We want to focus now on the situation $\sigma_\alpha(t)=\{E_\alpha(t)\}$, that is
when $P_\alpha(t)$ corresponds to an isolated
eigenvalue $E_\alpha(t)$, not necessarily simple, possibly associated with an
infinite dimensional spectral subspace. That means that
\bea\label{hamsingle}
H_\alpha(t)&=&E_\alpha(t)P_\alpha(t)+\overline{P_\alpha}(t)H_\alpha(t)\overline{P_\alpha}(t)\\
\label{res}(H_\alpha(t)-z)^{-1}&=&\frac{P_\alpha(t)}{E_\alpha(t)-z}+\overline{P_\alpha}(t)
(H_\alpha(t)-z)^{-1}\overline{P_\alpha}(t),
\eea
where the reduced resolvent
$\overline{P_\alpha}(t)(H_\alpha(t)-z)^{-1}\overline{P_\alpha}(t)$ is
holomorphic for all $z$'s inside the loop $\gamma$ of the definition (\ref{riesz}) if
$P_\alpha(t)$.

This case allows to distinguish nicely contributions
 from the ``dynamical phase'' and from the ``geometric phase'' in the usual
adiabatic language, in the periodic case. This comes as a simple
\begin{cor} \label{isolev}Assume $\sigma_\alpha(t)=\{E_\alpha(t)\}$, then
\be
P_\alpha(0)Q_\alpha(t)P_\alpha(0)=P_\alpha(0)\left(\frac{1}{\eps}\int_0^t
\partial_\alpha E_\alpha(s)ds
+iW_\alpha^{-1}(t)\partial_\alpha W_\alpha(t)\right)P_\alpha(0)+O(\eps).
\ee
\end{cor}
{\bf Remarks:}\\
i) We do not use periodicity in any of the variables yet.\\
ii) The form of the order zero term is similar to (\ref{alternate}),
which allows to interpret it as the geometrical charge transported
in the adiabatic process. This is supported by the fact that this term
is determined by the spectral projector $P_\alpha(t)$ only.\\
{\bf Proof:}
We will approximate $P^{(1)}_\alpha(t)
H^{(1)}_\alpha(t)$ by $E_\alpha(t)P^{(1)}_\alpha(t)$ by perturbation theory in $\eps$.
By means of the Riesz formula (\ref{riesz}) for $P^{(1)}_\alpha(t)$ we can write
\bea
P^{(1)}_\alpha(t)(H^{(1)}_\alpha(t)-E_\alpha(t))&=& -\frac{1}{2\pi i}
\int_{\gamma}(H^{(1)}_\alpha(t)-z)^{-1}(H^{(1)}_\alpha(t)-E_\alpha(t)) \, dz\nonumber\\
\quad \quad \quad &=&-\frac{1}{2\pi i}\int_{\gamma} (z-E_\alpha(t))
(H^{(1)}_\alpha(t)-z)^{-1}\, dz.
\eea
We used $\int_{\gamma}\un \, dz=0$.
Introducing the perturbed resolvent
(\ref{perres}), we get
\bea\label{singh}
&&P^{(1)}_\alpha(t)(H^{(1)}_\alpha(t)-E_\alpha(t))=
-\frac{1}{2\pi i}\int_{\gamma} (z-E_\alpha(t))(H_\alpha(t)-z)^{-1}\, dz\nonumber\\
&&-\frac{\eps}{2\pi i}\int_{\gamma} (z-E_\alpha(t))(H_\alpha(t)-z)^{-1}K_\alpha(t)
(H_\alpha(t)-z)^{-1} +O(\eps^2),
\eea
where the remainder keeps being of oder $\eps^2$ when differentiated. By making use of (\ref{res}) and the fact
that the reduced resolvent is analytic inside $\gamma$, one gets from Cauchy formula
that the first term of the right hand side is zero whereas the second yields
\be
-\frac{\eps}{2\pi i}\int_{\gamma} (z-E_\alpha(t))(H_\alpha(t)-z)^{-1}K_\alpha(t)
(H_\alpha(t)-z)^{-1}=\eps P_\alpha(t)K_\alpha(t)P_\alpha(t).
\ee
This term is zero due to (\ref{dk}), hence
\be
P^{(1)}_\alpha(t) H^{(1)}_\alpha(t)=E_\alpha(t)P^{(1)}_\alpha(t)+O(\eps^2),
\ee
where the remainder term can be differentiated. Therefore
\be
\partial_\alpha H^{(1)}_\alpha(t)=
\partial_\alpha (E_\alpha(t)P^{(1)}_\alpha(t)+\overline{P_\alpha}(t)
H^{(1)}_\alpha(t)\overline{P_\alpha}(t))+O(\eps^2)
\ee
and we get
\be
P^{(1)}_\alpha(t) \partial_\alpha H^{(1)}_\alpha(t)P^{(1)}_\alpha(t)=
 P^{(1)}_\alpha(t)\partial_\alpha E_\alpha(t)+O(\eps^2).
\ee
This allows to further simplify the first term in the expression
of Proposition \ref{prop}, making use of (\ref{perpro}) and (\ref{fut})
to get
\bea
&&P_\alpha(0)
\left(\frac{1}{\eps}\int_0^t
{\Phi^{(1)}_\alpha}^{-1}(s){W^{(1)}_\alpha}^{-1}(s)
P^{(1)}_\alpha(s) \partial_\alpha H^{(1)}_\alpha(s)
P^{(1)}_\alpha(s) W^{(1)}_\alpha(s)\Phi^{(1)}_\alpha(s)ds \right) P_\alpha(0)
\nonumber\\
&&= P_\alpha(0)
\frac{1}{\eps}\int_0^t \partial_\alpha E_\alpha(s)P^{(1)}_\alpha(0)\,
ds \, P_\alpha(0) +O(\eps)
\nonumber\\
&&=P_\alpha(0)\frac{1}{\eps}\int_0^t \partial_\alpha E_\alpha(s)\,
ds +O(\eps).
\eea
The last term in the expression of  Proposition \ref{prop} is dealt with
as follows. The condition (\ref{hamsingle}) implies
\be
\Phi_\alpha^{\pm 1}(t)P_\alpha(0)=
P_\alpha(0)e^{\mp \frac{i}{\eps}\int_{0}^tE_\alpha(s)\, ds}
\ee
so that we are left with
\be
P_\alpha(0)\left( \int_0^t W_\alpha^{-1}(s)
\partial_\alpha K_\alpha(s)
W_\alpha(s) ds\right) P_\alpha(0).
\ee
The argument leading form (\ref{defq}) to (\ref{alternate}) depends on
the differential equation satisfied by $U_\alpha (t)$ only, and thus
applies to $W_\alpha(t)$ as well, whose generator is $K_\alpha(t)$,
{\em mutatis mutandis}. This ends the
proof of the Corollary. \ep

\vspace{.3cm}

Further specializing to the periodic case
we get
\begin{cor}\label{evper}
Assume $\sigma_\alpha(t)=\{E_\alpha(t)\}$ and suppose $t\mapsto H_\alpha(t)$ is periodic in $t$, of period $1$. Then,\vspace{.2cm} \\
i) if  $E_\alpha(t)$ is non-degenerate
\be
\bra \ffi_\alpha(0)|Q_\alpha(1)\ffi_\alpha(0)\ket =\frac{1}{\eps}\int_0^1
\partial_\alpha E_\alpha(s)ds + \partial_\alpha \beta_\alpha+O(\eps),
\ee
where $\ffi_\alpha(0)$ is any normalized eigenvector at $t=0$ and
$\beta_\alpha$ is the corresponding geometric or Berry phase,\vspace{.2cm}\\
ii) if $E_\alpha(t)$ is degenerate and $\{\ffi_\alpha^{(r)}(0), | \ r\in \N\}$
denotes an orthonormal basis of $P_\alpha(0){\cal H}$, we have
\be
\bra \ffi_\alpha^{(r)}(0)|Q_\alpha(1)\ffi^{(s)}_\alpha(0)\ket =
\frac{1}{\eps}\int_0^1 \partial_\alpha E_\alpha(s)ds +
i\bra\ffi_\alpha^{(r)}(t) |\partial_\alpha\ffi_\alpha^{(s)}(t) \ket |^1_0+O(\eps),
\ee
where $\ffi_\alpha^{(r)}(t)=W_\alpha(t)\ffi_\alpha^{(r)}(0)$,
$r\in\N$,\vspace{.2cm}\\
iii) if $E_\alpha(t)$ is finitely degenerate, and if
$\{\psi_\alpha^{(r)}(t), | \ r\in \{1,2, \cdots, N\}\}$ denotes a $C^1$,
1-periodic orthonormal basis of  $P_\alpha(t){\cal H}$, we can write
\bea
&&\bra \ffi_\alpha^{(r)}(0)|Q_\alpha(1)\ffi^{(s)}_\alpha(0)\ket =
\frac{1}{\eps}\int_0^1 \partial_\alpha E_\alpha(s)ds \nonumber\\
&&+
i\left(\sum_{q,q'}\overline{B_\alpha^{q', r}}(1)B_\alpha^{q, s}(1)
\bra  \psi_\alpha^{(q')}(0)| \partial_\alpha \psi_\alpha^{(q)}(0) \ket
+\sum_q \overline{B_\alpha^{q, r}}(1)
\partial_\alpha B_\alpha^{q, s}(1)\right. \nonumber\\
&&\left.\phantom{\sum_{q,q'}}\hspace{-.3cm}-\bra\psi_\alpha^{(r)}(0) |
\partial_\alpha \psi_\alpha^{(s)}(0)\ket\right)+O(\eps),
\eea
where $B_\alpha(t)$ solves the ODE
\be
\dot{B}_\alpha(t)= \Gamma_\alpha(t) B_\alpha(t), \ \  B_\alpha(0)=\un ,
\ee
with $\Gamma_\alpha(t)$ defined by its matrix elements in the
basis $\{\psi_\alpha^{(r)}(0)\}_{r=1,\cdots, N}$
\be
\bra\psi_\alpha^{(s)}(0)| \Gamma_\alpha(t) \psi_\alpha^{(r)}(0)\ket
=-\bra \psi_\alpha^{(s)}(t) |\dot{\psi}_\alpha^{(r)}(t) \ket.
\ee
\end{cor}
{\bf Remarks:}\\
i) An explicit quantity for the geometric part of the charge transported
is always available in the non-degenerate case only, see (\ref{exbe}). In the
degenerate case, the geometric part is determined by the solution to
a (second order at least) ordinary differential equation.
No explicit solution is available in general and, moreover, the equation
is parameter free which forbids an asymptotic analysis. However, as
we explain below, there are special cases of interest in which an explicit
expression is available for this geometric contribution.  \\
ii) The third point is a mere restatement of the second one, making use
of an {\em a priori} time dependent basis of the eigenspace provided by
an independent spectral analysis.\\

\noindent
{\bf Proof:} To get the second statement, we compute
\bea
&&i\bra\ffi_\alpha^{(r)}(0) |W^{-1}_\alpha(t)(\partial_\alpha W_\alpha(t))
\ffi_\alpha^{(s)}(0) \ket=\\  \nonumber
&&i\bra\ffi_\alpha^{(r)}(0) |(W^{-1}_\alpha(t)(\partial_\alpha W_\alpha(t)
\ffi_\alpha^{(s)}(0))
-W^{-1}_\alpha(t) W_\alpha(t)\partial_\alpha
\ffi_\alpha^{(s)}(0)) \ket=\\ \nonumber
&&i\bra\ffi_\alpha^{(r)}(t) |\partial_\alpha\ffi_\alpha^{(s)}(t) \ket-
i\bra\ffi_\alpha^{(r)}(0) |\partial_\alpha\ffi_\alpha^{(s)}(0) \ket.
\eea
The first statement follows from
 $\ffi_\alpha^{(s)}(t)=\ffi_\alpha^{(r)}(t)=\ffi_\alpha(t)$ together with
the expression (\ref{berry}).\\
Finally, the third statement is proven as follows. Let us introduce
\be
\ffi_\alpha^{(r)}(t)=W_\alpha(t)\psi_\alpha^{(r)}(0) \ \ \mbox{and} \ \
\psi_\alpha^{(r)}(t)=
V_\alpha(t)\psi_\alpha^{(r)}(0), \ \ r=1, \cdots, N,
\ee
which defines the unitary $V_\alpha(t)$. The link
between these two bases will be made by means of the unitary operator
$B_\alpha(t)$ defined by
\be
B_\alpha(t)=V^{-1}_\alpha(t)W_\alpha(t).
\ee
By construction, $[B_\alpha(t),P_\alpha(0)]=0$ for any $t\in[0,1]$,
so that $[\dot{B}_\alpha(t),P_\alpha(0)]=0$ as well. We compute
\be
\dot{B}_\alpha(t)=B_\alpha(t)W^{-1}_\alpha(t)\dot{W}_\alpha(t)
+\dot{V}^{-1}_\alpha(t)V_\alpha(t)B_\alpha(t),
\ee
where the first term of the right hand side is zero due to (\ref{dk}).
Hence, introducing
\be
\Gamma_\alpha(t)=\dot{V}^{-1}_\alpha(t)V_\alpha(t)=
-V^{-1}_\alpha(t)\dot{V}_\alpha(t)
\ee
whose matrix elements in the basis
$\{\psi_\alpha^{(r)}(0)\}_{ r\in \{1,2, \cdots, N\}}$ read
\be
\bra\psi_\alpha^{(s)}(0)| \Gamma_\alpha(t) \psi_\alpha^{(r)}(0)\ket
=-\bra \psi_\alpha^{(s)}(t) |\dot{\psi}_\alpha^{(r)}(t) \ket,
\ee
we get that $B_\alpha(t)$ is indeed determined by the ODE
\be\label{defb}
\dot{B}_\alpha(t)= \Gamma_\alpha(t) B_\alpha(t), \ \  B_\alpha(0)=\un .
\ee
Writing $W_\alpha=V_\alpha B_\alpha$, we compute
\bea\label{aaa}
\bra \ffi_\alpha^{(r)}(t)|\partial_\alpha \ffi_\alpha^{(s)}(t)\ket
&=& \bra \ffi_\alpha^{(r)}(0)|\partial_\alpha \ffi_\alpha^{(s)}(0)\ket
\\ \nonumber
&& +\bra \psi_\alpha^{(r)}(0)|
B_\alpha^{-1}(t)V_\alpha^{-1}(t)(\partial_\alpha V_\alpha(t))
B_\alpha(t)  \psi_\alpha^{(s)}(0)\ket\\ \nonumber &&+
\bra \psi_\alpha^{(r)}(0)|
B_\alpha^{-1}(t)(\partial_\alpha B_\alpha(t))
 \psi_\alpha^{(s)}(0)\ket .
\eea
With the short hand $B_\alpha^{q, r}(t)=\bra \psi_\alpha^{(q)}(0)|
B_\alpha \psi_\alpha^{(r)}(0) \ket $, we have
\bea
B_\alpha(t)  \psi_\alpha^{(s)}(0)&=&\sum_q \psi_\alpha^{(q)}(0)
B_\alpha^{q, s}(t)\\
V_\alpha(t)B_\alpha(t)  \psi_\alpha^{(s)}(0)&=&\sum_q \psi_\alpha^{(q)}(t)
B_\alpha^{q, s}(t),
\eea
and
\bea
(\partial_\alpha V_\alpha(t))B_\alpha(t)  \psi_\alpha^{(s)}(0)
&=&\sum_q B_\alpha^{q, s}(t)((\partial_\alpha V_\alpha(t) \psi_\alpha^{(q)}(0))-
 V_\alpha(t) \partial_\alpha\psi_\alpha^{(q)}(0))\nonumber\\
&=&\sum_q B_\alpha^{q, s}(t)(\partial_\alpha \psi_\alpha^{(q)}(t)
- V_\alpha(t) \partial_\alpha\psi_\alpha^{(q)}(0))\nonumber\\
 (\partial_\alpha B_\alpha(t))  \psi_\alpha^{(s)}(0)&=&\sum_q
\partial_\alpha( B_\alpha^{q, s}(t)\psi_\alpha^{(q)}(0))-
B_\alpha(t)\partial_\alpha\psi_\alpha^{(s)}(0)
\eea
Inserting these expressions in (\ref{aaa}), we get
\bea
&&  \bra \ffi_\alpha^{(r)}(t)|\partial_\alpha \ffi_\alpha^{(s)}(t)\ket
-\bra \ffi_\alpha^{(r)}(0)|\partial_\alpha \ffi_\alpha^{(s)}(0)\ket
\nonumber\\
&&=
\sum_{q,q'}\overline{B_\alpha^{q', r}}(t)B_\alpha^{q, s}(t)
\bra  \psi_\alpha^{(q')}(t)| \partial_\alpha \psi_\alpha^{(q)}(t) \ket
+\sum_q \overline{B_\alpha^{q, r}}(t)
\partial_\alpha B_\alpha^{q, s}(t) \nonumber\\
&&-\bra\psi_\alpha^{(r)}(0) |
\partial_\alpha \psi_\alpha^{(s)}(0)\ket,
\eea
which yields the result.
\ep\\

{\bf Note:} The operator $B_\alpha(t)$ and its generator $\Gamma_\alpha(t)$
depend of course on the choice of orthonormal basis $\{\psi_\alpha^{(r)}(t)\}_{r=1,\cdots, N}$.
It is not difficult to check that if one makes another choice of orthonormal basis
 $\{\chi_\alpha^{(r)}(t)\}_{r=1,\cdots, N}$ such that
$\chi_\alpha^{(r)}(t)=S_\alpha(t)\psi_\alpha^{(r)}(0)$, then the corresponding
generator denoted by $\Sigma_\alpha(t)$ is related to the previous one by
means of $C_\alpha(t)=S_\alpha^{-1}(t)V_\alpha(t)$ according to
\bea
\Sigma_\alpha(t)&=&
C_\alpha(t)\Gamma_\alpha(t)C^{-1}_\alpha(t)+\dot C_\alpha(t)C^{-1}_\alpha(t).
\eea

\section{Example}

We consider here an explicit class of Hamiltonians which, on the one hand,
display permanent degeneracies, and, on the other hand, allow in some cases
for explicit computations. Moreover, the physical situation considered in
\cite{bfhj} is governed by a Hamiltonian of this class.\\

Let $\{z_1, z_2, \cdots, z_n\}$ be a set of $n$ complex numbers, which we
denote  by the vector $z=(z_1,\cdots, z_n)^T\in \C^n$, and let $E\in\R$.
Let us denote the standard scalar product in $\C^n$ by $\bra \,\cdot\, |\,\cdot \,\ket$.
We consider the self-adjoint Hamiltonian
\be\label{rank}
H(z)=\begin{pmatrix}{ E  & \bar{z_1}  & \cdots & \bar{z_n}\cr
                z_1 & 0 & \cdots & 0 \cr
                 \vdots & \vdots  & \ddots & \vdots \cr
                 z_n & 0 & \cdots & 0}
 \end{pmatrix}\equiv \begin{pmatrix}{ E  & \bra z|\cr
                |z\ket  & 0 }
 \end{pmatrix} \ \ \mbox{on} \ \ \C^{n+1}\simeq \C\oplus\C^n,
\ee
relative to the canonical basis $\{e_0, e_1, \cdots, e_n\}$ of
$\C^{n+1}$. We made $z$ explicit in the notation because these
parameters will become time-dependent below.

If $z\neq 0$, the rank of $H(z)$ is equal to two, so that
its kernel if of dimension $n-1$, for any value of the parameters.
If $z=0$, the kernel of $H(0)$ is of dimension $n$. Actually, it is
easy to see that
\be
\sigma(H(z))=\left\{\frac12(E-\sqrt{E^2+4\|z\|^2}), 0, \frac12(E+\sqrt{E^2+4\|z\|^2})\right\},
\ee
where $\|z\|^2=\sum_{j=1}^n|z_j|^2$, and where the eigenspace corresponding to the
$(n-1)$-fold degenerate eigenvalue $0$ is given by
\be
\ker (H(z))=\left\{ \begin{pmatrix}{a_0 \cr a_1 \cr \vdots \cr a_n}\end{pmatrix}\in\C^{n+1}
 \ \mbox{s.t.}
\ a_0=0 \ \mbox{and}\ \sum_{j=1}^n \bar{z_j}a_j=0\right\}.
\ee
 We
can rewrite with $a=(a_1,\cdots, a_n)^T\in \C^n$
\be
\ker (H(z))=0\oplus \left\{a\in\C^{n} \
\mbox{s.t.}\ \bra z | a\ket =0\right\}= 0\oplus z^\perp,
\ee
where $z^\perp$ denotes the orthogonal of the vector $z\in\C^n$. It is now easy
to express the projector $P(z)$ on the degenerate spectral subspace
$\ker (H(z))$ in $\C^{n+1}$. Let $\hat z=z/\|z\|\in\C^n$ and $|\hat z\ket\bra \hat z|$
be the projector on the vector $\hat z$ in $\C^n$. Hence,
\be
P_\perp(z)=\un_{\C^n}-|\hat z\ket\bra \hat z|
\ee
is the projector on $z^\perp$ in $\C^n$. Thus, expressed in block diagonal form in
$\C^{n+1}\simeq \C\oplus\C^n$, we can write $P(z)$ as
\be
P(z)=\begin{pmatrix}{0 & 0 \cr 0 & P_\perp(z)}\end{pmatrix}.
\ee
Hence, with the same notations, $\overline{P}(z)$  can be written as
\be
\overline{P}(z)=\begin{pmatrix}{1 & 0 \cr 0 & |\hat z\ket\bra \hat z|}\end{pmatrix},
\ee
so that the range of $ \overline{P}(z)$ is generated by the orthonormal basis
\be\label{rpb}
\mbox{Ran}\, \overline{P}(z)=\mbox{span} \left\{\begin{pmatrix}{1  \cr 0}\end{pmatrix},
\begin{pmatrix}{0  \cr \hat z}\end{pmatrix}\right\}\equiv \mbox{span} \left\{
e_0, \tilde z\right\}.
\ee

Let as assume now that $z=z(t)$ is time-dependent, in such a way that
$[0,1]\ni t\mapsto z(t)\in \C^n$ is $C^3$. By changing the phase of
$z(t)$ if necessary, we can assume
\be\label{ber}
\bra \hat z(t) |\dot {\hat z}(t)\ket=\bra \tilde z(t) |\dot {\tilde z}(t)\ket\equiv 0.
\ee
It is now straightforward to check that the parallel transport operator
$W(t)$ is generated by the self-adjoint operator $K(t)=i[\dot P(z(t)), P(z(t))]$, with
\be
K(t)=i(|\dot {\tilde z}(t)\ket\bra \tilde z(t) |-| \tilde z(t) \ket\bra \dot {\tilde z}(t) |).
\ee
Recall that $\tilde z=(0  , \hat z)^T$ is a normalized a vector
of $\C^{n+1}$.
Note that condition (\ref{ber}) is equivalent to
saying
\be
\tilde z(t)=W(t)\tilde z(0) \ \ \mbox{and} \ \
\begin{pmatrix}{1  \cr 0}\end{pmatrix}=W(t)\begin{pmatrix}{1  \cr 0}\end{pmatrix}.
\ee

Thus the determination of $W$ restricted to $\overline{P}$ is complete. With these preliminaries behind us, we can turn to the
interesting task from our point of view, {\it i.e.} the determination of $W$ restricted to
$P$. From
(\ref{rpb}) above, it is clear that we can restrict attention to $\C^n\simeq e_0^\perp$, where
$\mbox{Ran}\, \overline{P}(z)\cup e_0^\perp\simeq \C\hat z$.\\

Let $\{\ffi_1, \ffi_2, \cdots, \ffi_{n-1}\}$ be an orthonormal basis in $\C^n$
of $\hat z(0)^\perp$. Then, using the same notation for $\ffi_j\in \C^n$ and
$(0, \ffi_j)^T\in\C^{n+1}$, we have for any $j=1,\cdots, n$, and any $t\in [0,1]$,
\be\label{partra}
\ffi_j(t)=W(t)\ffi_j\in \C^{n+1} \ \Leftrightarrow \ \left\{\begin{array}{l}\dot \ffi_j(t)=-
| \hat z(t) \ket\bra \dot {\hat z}(t) |\ffi_j(t)\ket \in \C^n
\\ \ffi_j(0)=\ffi_j\in\C^n.\end{array}\right.
\ee

Actually, computing  the parallel transport operator $W(t)$ restricted to
$\ker(H(z(t))$ for the model (\ref{rank}) where $z(t)\in\C^n$ is given, amounts
to determining $n-1$ vectors
$\ffi_j(t)$ in $\C^n$ such that for all $j, k=\{1,\cdots, n-1\}^2$ and all $t\in [0,1]$,
\bea
\bra \ffi_j(t)|\hat z(t)\ket&\equiv& 0\\
\bra \ffi_j(t)|\ffi_k(t))\ket&\equiv &\delta_{jk}\\
\bra \ffi_j(t)|\dot\ffi_k(t))\ket&\equiv &0.
\eea
Indeed, if (\ref{partra}) is satisfied, the conditions above are met. Conversely,
if the first two conditions above are satisfied, we get that
$\{\hat z(t), \ffi_1(t), \cdots, \ffi_{n-1}\}$ form an orthonormal basis, for all $t$'s.
Moreover, the third  condition implies  that $\dot \ffi_j=c_0(t)\hat z(t)$, for some
coeffcient $c_0(t)\in\C$. Differentiation of $\bra \ffi_j(t)|\hat z(t)\ket\equiv 0$ yields
$a_0(t)=-\bra\dot {\hat z}(t)|\ffi_j(t)\ket$, so that equations (\ref{partra}) are true.\\

Eventhough the generator of $W$ restricted to $P(z(t))$ is rather simple, these equations
cannot be explicitely integrated in general. We present some special cases of interest
which allow for explicit formulas.

\subsection{Special case}

We consider here a special case for $n=3$ that is of interest for the physics of charge pumping, \cite{bfhj}. Let us consider the Hamiltonian
\be
H_0(z_0, z_1, z_3)=
\pmatrix
{E & \bar{z_0}& \bar{z_1} & \bar{z_3}\cr
 z_0 & 0 & 0 & 0 \cr
z_1  & 0 & 0 & 0 \cr
z_2  & 0 & 0 & 0 }
\ee
in the canonical basis.
We assume that
\be\label{cond}
|z_1|^2+|z_2|^2>0,
\ee so that
a set of normalized eigenvectors corresponding to the degenerate subspace
of energy zero is given by
\bea
\overline{|\psi_{1}\ket}&=&N_{1}(z_2\,e_2 - z_1\,e_3) \\
\overline{|\psi_{2}\ket}&=&N_{1}\left[ (z_1^2+z_2^2)\,e_1 - z_0
(z_1\,e_2 + z_2\,e_3)\right],
\eea
with
\bea
N_{1}&=&1/\sqrt{|z_1|^2+|z_2|^2}\\
N_{2}&=&1/\sqrt{|z_1^2+z_2^2|^2+|z_0|^2(|z_1|^2+|z_2|^2)}.
\eea

We now compute the differentials of these eigenvectors, in order to get
the generator of the non-abelian transformation. At this level, we allow all
parameters to vary, with the condition that
(\ref{cond}) holds. Straightforward computations yield the (negative of) the matrix elements of the matrix $\Gamma_\psi$, with respect to this instantaneous
basis of eigenvectors of $\ker H$
\bea\label{gampsi}
\bra \psi_{1}| d  \psi_{1}\ket&=&iN_1^2\Im (z_1d\bar{z_1}+z_2d\bar{z_2})\\ \nonumber
\bra \psi_{2}| d  \psi_{2}\ket&=&i N_2^2 \Im\left\{2(z_1^2+z_2^2)(\bar{z_1}d\bar{z_1}+\bar{z_2}d\bar{z_2})+|z_0|^2({z_1}d\bar{z_1}+{z_2}d\bar{z_2})+
 (|z_1|^2+|z_2|^2)z_0d\bar{z_0}  \right\}
\\  \nonumber
\bra \psi_{2}| d  \psi_{1}\ket&=&N_1N_2z_0(z_2d\bar{z_1}-z_1d\bar{z_2})
\\  \nonumber
\bra \psi_{1}| d  \psi_{2}\ket&=&-\overline{\bra \psi_{2}| d  \psi_{1}\ket}.
\eea
We can simplify some more this matrix by passing to the time-dependent basis
\be
\chi_j(t)=\psi_j(\gamma(t))e^{i\beta_j(t)}, \ \ \beta_j(t)=i\int_0^t\bra\psi_j |d\psi_j\ket(\gamma(s)) ds,
\ee
where the integral is taken along a path $[0,1]\ni t\mapsto \gamma(t)$ in the parameters
space. The matrix $\Gamma_{\chi}$ corresponding to the basis $\{\chi_1, \chi_2\}$ of eigenvectors of $\ker H$  now reads
\be
\Gamma_{\chi}=\pmatrix{0 & -e^{i(\beta_2-\beta_1)}\bra \psi_1|d\psi_2\ket\cr
e^{-i(\beta_2-\beta_1)}\overline{\bra \psi_1|d\psi_2\ket}  & 0}.
\ee
Setting
\be
x(t)=e^{-i(\beta_2(t)-\beta_1(t))}\overline{\bra \psi_1|d\psi_2\ket}(\gamma(t)),
\ee
so that
\be\label{easy}
\Gamma_{\chi}(t)=\pmatrix{0 & -\overline{x(t)}\cr x(t) & 0},
\ee
we have to solve the ODE $\dot B(t)=\Gamma_{\chi}(t) B(t)$, see (\ref{defb}), to
determine $W(t)$.
In general, no explicit solution to (\ref{defb}) with such a matrix can be obtained.

However, in case $x(t)=\rho(t)e^{i\vartheta}$, where $\vartheta$ is constant in time, $\Gamma_{\chi}(t)=\rho(t)M$, where $M=\pmatrix{0 & -e^{-i\vartheta}\cr e^{i\vartheta} & 0}$ and
$B(t)$ is explicitely given by
\be\label{exb}
B(t)=e^{\int_0^t \rho(s)ds M}=\pmatrix{\cos(\int_0^t \rho(s)ds)& -\sin(\int_0^t \rho(s)ds)e^{-i\vartheta}\cr\sin(\int_0^t \rho(s)ds)e^{i\vartheta}& \cos(\int_0^t \rho(s)ds)}.
\ee
We consider below a case of this type, which allows to determine explicitely the geometric part of the transported charge over a period. Moreover, we express the
geometric content if the parallel transport within the {\em permanently degenerate} kernel of $H$ as a solid angle in the space of parameters, in a similar fashion to what is done for the Berry phase, in case of {\em non-degenerate} eigenvalues.

Let us assume that
\be
z_j=e^{i\theta_j}r_j \ \ \mbox{and} \ \ \ dz_j=e^{i\theta_j}dr_j,
\ee
that is, only the moduli of the complex numbers $z_j$ vary with time. Plugging
this into (\ref{gampsi}) yields
\bea
\bra \psi_{1}| d  \psi_{1}\ket&=&0\\ \nonumber
\bra \psi_{2}| d  \psi_{2}\ket&=& 2iN_2^2\sin(2(\theta_1-\theta_2))r_1r_2(r_1dr_2-r_2dr_1)\\ \nonumber
\bra \psi_{2}| d  \psi_{1}\ket&=& N_1N_2r_0e^{i\theta_0}
\left(e^{-i(\theta_1- \theta_2)}r_2dr_1-e^{i(\theta_1- \theta_2)}r_2dr_2 \right)
\eea
with
\be
N_1=\frac{1}{\sqrt{r_1^2+r_2^2}}, \ \ N_2=\frac{1}{\sqrt{(r_1^2+r_2^2)r_0^2+(r_1^4+r_2^4+2r_1^2r_2^2\cos(2(\theta_1- \theta_2)))}}.
\ee
Further assuming
\be
\theta_1=\theta_2=0,
\ee
we finally get
\be
\Gamma_\psi=\frac{r_0(r_1dr_2-r_2dr_1)}{(r_1^2+r_2^2)\sqrt{r_0^2+r_1^2+r_2^2}}\pmatrix{0& -e^{-i\theta_0}\cr e^{i\theta_0} & 0},
\ee
which is of the form (\ref{easy}). The argument of the sines and cosines in
(\ref{exb}) after a period caracterized by a loop $\gamma$ in the space of
parameters is denoted by
\be
\Omega=\int_\gamma \frac{-r_0(r_1dr_2-r_2dr_1)}{(r_1^2+r_2^2)\sqrt{r_0^2+r_1^2+r_2^2}}
\ee
so that
\be
B(1)=\pmatrix{\cos(\Omega)& \sin(\Omega)e^{-i\theta_0}\cr-\sin(\Omega)e^{i\theta_0}& \cos(\Omega)}.
\ee

Similarly, if $z_1$ and $z_2$ are as above and $z_0=t_1e^{i\theta_0}+t_2$ with $t_1, t_2$ real and $dr_1=dr_2=d\theta_0\equiv 0$, with $r_1^2+r_2^2>0$, we have
\be
\Gamma_\psi=  i\sin(\theta_0) (t_1dt_2-t_2dt_1)/(t_1^2+t_2^2+2t_1t_2\cos(\theta_0))
\pmatrix{0& 0\cr 0& 1}.
\ee

\subsection{Geometric interpretation of $\Omega$}

The explicit computation of $\Omega$ possesses
a nice geometric interpretation, see (\ref{geom}), as we now explain.

For notational convenience, let us introduce cartesian coordinates
$(x,y,z)=(r_0, r_1, r_2)$. At the end of the loop $\gamma$,
we have
\bea
\Gamma&=&\int_\gamma\frac{-z(xdy-ydx)}{(x^2+y^2)\sqrt{x^2+y^2+z^2}}\\
&=&\int_\gamma\frac{-z}{(x^2+y^2)\sqrt{x^2+y^2+z^2}}\pmatrix{-y\cr x \cr 0}
\cdot \pmatrix{dx\cr dy\cr dz}.
\eea
Applying Stokes' Theorem, we can replace the above curvilinear integral by the flux
of the curl of the corresponding vector through
any surface $\Sigma$ such that $\partial \Sigma=\gamma$, oriented
consistently with the orientation of $\gamma$.\\
We compute
\bea
\vec{\mbox{curl}}\ \frac{-z}{(x^2+y^2)\sqrt{x^2+y^2+z^2}}\pmatrix{-y\cr x \cr 0}&=&
(x^2+y^2+z^2)^{-3/2}\pmatrix{x\cr y \cr z}\\
&=&\frac{\vec r}{r^3},
\eea
where
$
\vec r=\pmatrix{x\cr y \cr z}$ and
$r=\sqrt{x^2+y^2+z^2}$.
Hence we can write,
\be\label{int}
\int_\gamma\frac{-z(xdy-ydx)}{(x^2+y^2)\sqrt{x^2+y^2+z^2}}=\int_{\Sigma}\frac{\vec r}{r^3}
\cdot d\vec \sigma.
\ee
Consider the projection $\widehat \gamma$ of the loop $\gamma$ on the sphere
$\mathbb S^2$  described by the unit vector $\hat r=\vec r/r$ along $\gamma$, and
define $\widehat \Sigma \subset \mathbb S^2$ such that $\partial \widehat
\Sigma=\widehat \gamma$. Now, we can choose for $\Sigma$ the surface which
coincides with $\widehat \Sigma$, and joins $\widehat \gamma$ and $\gamma$
along rays parallel to the unit vector. Since the flux of $\frac{\vec r}{r^3}$
through the latter portions of $\Sigma$ is zero, we finally get
\bea
\int_\gamma\frac{-z(xdy-ydx)}{(x^2+y^2)\sqrt{x^2+y^2+z^2}}&=&
\int_{\widehat \Sigma}\frac{\vec r}{r^3}
\cdot d\vec \sigma\\
&=&\int_{\widehat \Sigma}d\omega,
\eea
where we used the fact that on $\mathbb S^2$, $ d\vec \sigma=d\omega \hat r$,
with $d\omega $ the differential of the solid angle.\\

Therefore, we have obtained
\be\label{geom}
\Omega = \Omega(\widehat \Sigma)
\ee
where $\Omega(\widehat \Sigma)$
is the oriented solid angle described by $\gamma$ through $\mathbb S^2$. If $\widehat \gamma$ is oriented positively, $\Omega(\widehat \Sigma)\geq 0$, and,
in any case, $0\leq |\Omega(\widehat \Sigma)|\leq 4\pi$.


\newpage

\begin{thebibliography}{9}

%
\bibitem{aunola03}
  M. Aunola and J. J. Toppari, Phys. Rev. B {\bf 68}, 020502 (2003).

\bibitem{ae} J.E. Avron and A. Elgart,
Adiabatic Theorem without a Gap Condition,
{\em Commun. Math. Phys.} {\bf  203}, 445-463, (1999).

\bibitem{aegss} J.E. Avron, A. Elgart, G.M. Graf , L. Sadun and K. Schnee,
Adiabatic charge pumping in open quantum systems  {\em Comm. Pure and App. Math.} {\bf 57}, 528-561 (2004)

\bibitem{asy} J.E. Avron, R. Seiler and L.G. Yaffe, Adiabatic theorems
and applications to the quantum Hall effect,
{\it Commun. Math. Phys.} {\bf 110}, 33-49 (1987).

\bibitem{as} J.E. Avron, R. Seiler, Quantization of the Hall conductance
for general multiparticle Schr\"odinger Hamiltonian, {\em Phys. Rev. Lett.}
{\bf 54}, 259-262, (1985)

\bibitem{be} M.V. Berry, Quantal phase factors accompanying adiabatic changes,
{\em Proc. R. Soc. Lond. A} {\bf 392}, 45(1984)

\bibitem{bohm}
  \emph{The geometric Phase in quantum systems},
  A. Bohm {\em et al.} Eds., (Springer-Verlag, Berlin, 2003).

\bibitem{bf} M. Born, V. Fock, Beweis des Adiabatensatzes, {\em Zeit. f. Physik}
{\bf 51}, 165-180, (1928).

\bibitem{bornemann} F. Bornemann, ``Homogeneization in Time of Singularly Perturbed Mechanical
Systems'', {\em Lecture Notes in Mathematics} {\bf 1687}, Springer, Heidelberg, 1998.

\bibitem{bfhj} V. Brosco, R. Fazio, F. Hekking, A. Joye, Non-abelian
superconducting pumps, {\em Phys. Rev. Lett.} ,  {\bf 100}, 027002, (2008) {\em Preprint}

\bibitem{brow} P.W. Brouwer, Scattering approach to parametric pumping, {\em Phys. Rev. B} {\bf 58}, R10135 (1998)

\bibitem{btp} M. B\"uttiker, H. Thomas, A. Pr\^etre, Current partition in multi-probeconductors in the presence of slowly oscillating external potentials. {\em Z. Phys. B} {\bf 94} 133Ð137 (1994).

\bibitem{fhp} R. Fazio, F.W. J.
Hekking, J. P. Pekola, Phys. Rev. B {\bf 68 }, 0545410 (2003).
%
\bibitem{governale}
  M. Governale, F. Taddei, R. Fazio, and F. W. J. Hekking,


\bibitem{hjbham} G. Hagedorn, A. Joye, Recent Results on Non-Adiabatic Transitions
in Quantum Mechanics, In:  Recent Advances in Differential Equations and Mathematical Physics.
{\it AMS Contemporary Mathematics Series },  {\bf 412}, Ed. by N. Chernov, Y. Karpeshina, I. Knowles, R. Lewis, and R. Weikard. p. 183--198 2006.

\bibitem{thesis} A. Joye, ``Geometrical and mathematical aspects of the adiabatic
theorem in quantum mechanics'', EPFL thesis No 1022, 1992.
{\tt http://biblion.epfl.ch/EPFL/theses/1992/1022/EPFL\_TH1022.pdf}

\bibitem{joye} A. Joye, " General Adiabatic Evolution with a Gap Condition ",
{\it Commun. Math. Phys.} {\bf 275} 139-162 (2007).

\bibitem{jp2} A. Joye and C.-E. Pfister,  Superadiabatic evolution and
adiabatic transition probability
between two non-degenerate levels isolated in the spectrum,
{\it J. Math. Phys.} {\bf 34}, 454-479 (1993).

\bibitem{k1} T. Kato, On the adiabatic theorem of quantum mechanics,
{\em J. Phys. Soc. Japan}, {\bf 5}, 435-439, 1950.

\bibitem{k} T. Kato,  ``Perturbation Theory for Linear Operators'', Springer,
1980.
%

\bibitem{lang} S. Lang, ``Real Analysis'', Addison-Wesley, 1973.

\bibitem{mes} A. Messiah, "Quantum Mechanics", Dover Publications Inc., 2000.

\bibitem{n1} G. Nenciu, On the adiabatic theorem of quantum mechanics,
{\it J. Phys. A} {\bf 13} L15-L18 (1980).
%
\bibitem{n2} G. Nenciu, Adiabatic theorems and spectral concentration,
{\it Commun. Math. Phys.} {\bf 82} 121-135 (1981).

\bibitem{n3} G. Nenciu, Asymptotic invariant subspaces, adiabatic theorems
 and block diagonalisation, in
{\it Recent developments in quantum mechanics (Poiana Bra\c sov, 1989)},
{\it Math. Phys. Stud.}, {\bf 12}, 133-149,
Kluwer Acad. Publ., Dordrecht, 1991.

\bibitem{nt} Q. Niu, D.J. Thouless, Quantised adiabatic charge transport in the presence of substrate
disorder and many body interactions, {\it J. Phys. A}  {\bf 17}, 30-49 (1984)



\bibitem{leone}R. Leone and L. Levy, Phys. Rev. Lett. {\bf 100}, 117001 (2008).
%
  Phys. Rev. Lett. {\bf 95} 256801 (2005).

\bibitem{lidar}  M. S. Sarandy, D. A. Lidar :
Adiabatic approximation in open quantum systems,
{\em Phys. Rev. A} {\bf 71}, 012331 (2005).


%
\bibitem{mottonen06}M. M\"ott\"onen {\sl et al.}, Phys. Rev. B {\bf 73}, 214523 (2006).
%
\bibitem{mottonen08} M. M\"ott\"onen, J. J. Vartiainen, and J. P. Pekola, Phys. Rev. Lett. {\bf 100}, 177201 (2008).

\bibitem{pekola}
  J.P. Pekola, J.J. Toppari, M. Aunola, M.T. Savolainen,
  and D.V. Averin, Phys. Rev. B {\bf 60}, R9931 (1999).



\bibitem{simon}	B. Simon, Holonomy, the Quantum Adiabatic Theorem, and Berry's Phase, Phys. Rev. Lett. {\bf 51}, 2167 (1983).
%


\bibitem{t} S. Teufel ``Adiabatic perturbation theory in quantum dynamics'',
{\em Lecture Notes in Mathematics} {\bf 1821},
Springer-Verlag, Berlin, Heidelberg, New York (2003).


\bibitem{wilczek} F. Wilczek and A. Zee,	
Appearance of Gauge Structure in Simple Dynamical Systems, Phys. Rev. Lett. {\bf 52}, 2111 (1984).
%

\end{thebibliography}
\end{document}